\documentclass[sigconf]{acmart}

\usepackage{multirow}

\acmConference[PACT '22]{PACT '22:  International Conference on
Parallel Architectures and Compilation Techniques (PACT)}{October 10--12, 2022}{Chicago,IL}
\acmBooktitle{PACT '22: International Conference on
Parallel Architectures and Compilation Techniques (PACT), October 10--12, 2022, Chicago, IL}
\acmPrice{15.00}
\acmISBN{978-1-4503-XXXX-X/18/06}

\title{POAS: A high-performance scheduling framework for exploiting Accelerator Level Parallelism}
\author{Pablo Antonio Martínez}
\email{pabloantonio.martinezs@um.es}
\orcid{0000-0002-4391-2451}
\author{Gregorio Bernabé}
\email{gbernabe@um.es}
\orcid{0000-0002-7265-3508}
\author{José Manuel García}
\email{jmgarcia@um.es}
\orcid{0000-0002-6388-2835}
\affiliation{%
  \institution{University of Murcia}
  \country{Spain}
  \city{Murcia}
}
\date{April 2022}

\begin{document}

\begin{abstract}
Heterogeneous computing is becoming mainstream in all scopes.
This new era in computer architecture brings a new paradigm called Accelerator Level Parallelism (ALP).
In ALP, accelerators are used concurrently to provide unprecedented levels of performance and energy efficiency.
To reach that, there are many problems to be solved, one of the most challenging being co-execution.

This paper develops a scheduling framework called POAS, a general method for providing co-execution to generic applications.
Unlike other scheduling approaches, POAS does not directly schedule applications.
Instead, it is a generic model that transforms any application to make it suitable for co-execution, so that it can be executed in ALP environments.
Our proposal is composed of four differentiated steps: predict, optimize, adapt and schedule.
During these phases, different modifications are implemented in the application to make it suitable to be executed in ALP environments.
In this work we also apply our framework to a matrix multiplication case study, outlining the critical and most important steps to port the application with POAS.

We evaluate our POAS-based implementation for matrix multiplication on a CPU/GPU/XPU environment using CPU cores, CUDA cores and tensor cores (XPU).
Our experiments prove that co-execution in the studied scenario can benefit from ALP, yielding speedups of up to 45\% with respect to using only one accelerator.
The proven flexibility and potential of POAS make it an excellent candidate to reach ALP in future computer systems.
\end{abstract}




\settopmatter{printfolios=true}

\maketitle

\section{Introduction}
In recent years, it has been demonstrated that CPUs are less efficient regarding power consumption and performance compared to accelerators~\cite{acmacc}, which provide high performance that comes from parallelism enabled by specialization.
After the end of Moore's law~\cite{endmoore}, accelerators seem to be one of the few ways to keep improving the performance and efficiency of computing hardware.
Therefore, computer architecture is evolving to be heterogeneous, and to use different accelerators to accomplish different tasks, instead of relying on the CPU for all of them.
That is why the concept of using a general-purpose processor for everything is losing force over time, and computing has started the transition from being general-purpose toward specialization~\cite{acmgpt}.
In fact, it is said that this paradigm change is opening a new golden age for computer architecture~\cite{acmgoldenage}.

As computer science has evolved, many different paradigms have appeared to boost computers performance.
All of them (ILP, TLP and DLP) have been critical advances in computer architecture.
With such, there have been great opportunities to improve the CPU's performance.
Those opportunities were supported by the ability to keep increasing the transistor count inside the chips.
But after the end of Moore's law, similar techniques may not be appropriate since there is no physical space to implement them.
Since then, computer architecture is transcending to what some authors~\cite{acmalp} call the next computer architecture paradigm; Accelerator Level Parallelism (ALP).

Actually, this transformation from a general-purpose to a heterogeneous, specialized view started years ago.
Alternative architectures with lower overhead than the CPUs have been proposed, like Graphics Processing Units (GPUs).
The GPU is the most popular architecture, being more specialized than the CPU, but still more generic than other architectures.
Field-programmable gate arrays (FPGAs) play another crucial role towards specialization, as they allow solving specific problems directly in hardware in a programmatic way.
However, since accelerators are targeted to specific task/s, the majority of them are useful only for a particular domain.
They are used in domains like machine learning, where many accelerators exist, like the Tensor Processing Unit (TPU), the Neural Processing Unit (NPU), etc.
Other notable accelerators include Image Signal Processor (ISP), Digital Signal Processor (DSP), or video encoder/decoders.
In the end, many of these accelerators are often included in System on a Chip (SoC) that nowadays power our smartphones~\cite{anandtechsnapd8}, but also laptops~\cite{anandtechm1} and workstations~\cite{anandtechadl}.
Actually, SoCs are the first manifestation of ALP, as they include many accelerators that can be used concurrently, thus providing ALP~\cite{acmalp}.

Using many hardware devices concurrently is usually referred to as co-execution, and is one of the different goals that have to be fulfilled to achieve ALP.
The idea is to use many accelerators at the same time, similarly to how ILP concurrently employs multiple functional units.
Allowing co-execution is challenging since the software needs to divide the work into parts and schedule them among different devices.
This scheduling may pursue different objectives, like minimizing the execution time or the energy consumption.
In either case, achieving so depends heavily on the target hardware platform.
Therefore, many approaches are needed to cover all the possible scheduling domains.

This paper presents POAS (Predict, Optimize, Adapt and Schedule), a framework for adapting any workload to be executed concurrently on multiple accelerators.
The presented framework can be focused on minimizing the execution time (high-performance) or minimizing the energy consumption (energy efficiency).
The framework is divided into four general steps.
The first one, predict, consists of developing a solid model that predicts the execution time of the CPU and the accelerators, as well as the memory cost to copy the data between the CPU and the accelerators.
In the optimization step, the performance prediction model is used to build a constraint satisfaction problem (CSP).
The problem is then optimized to find the values such that the objective function is minimal.
Lastly, the results given by the solver may need to be adapted to the specific problem so that they can be used to schedule the workload.

To demonstrate how POAS works, this paper applies our method to a high-performance scenario like matrix multiplication.
First, the general method is specialized for the general matrix multiplication domain.
As target accelerators, we consider NVIDIA GPUs, both ones that use ordinary CUDA cores, as well as those that provide hardware acceleration for matrix multiplications using tensor cores (XPUs)~\cite{voltauarch}.
Therefore, ALP is achieved by exploiting the power of the host CPU, the GPU (or GPUs) and the XPU (or XPUs), which we refer to the GPUs with tensor cores.
The model is implemented and evaluated in two HPC environments with different performance capabilities, both having a multi-core CPU, a GPU and an XPU.
The results highlight the flexibility and great performance that the proposed method provides.

To the best of our knowledge, this work is the first proposal to define a generic scheduling framework for heterogeneous platforms to allow co-execution that can be adapted to particular domains. 
While it has the disadvantage of requiring extra work, it has the potential to provide better results, which is a trade-off that might be interesting in many different scenarios.

The main contributions of this paper are the following:
\begin{itemize}
\item Defines a framework for exploiting Accelerator Level Parallelism (ALP) through co-execution. 
The model consists of a four-step decomposition of the scheduling problem and is applicable in many applications.
\item Details how the proposed framework can be applied to a real-world application like matrix multiplication and implements it for CPU, GPUs and XPUs environments.
\item Presents an experimental evaluation of the proposed framework applied to the matrix multiplication domain, showing its performance gain in a real-world scenario, yielding speedups of up to 45\% with respect to using only one accelerator.
\end{itemize}

The rest of the paper is organized as follows. Section~\ref{sec:background} presents the background in scheduling techniques, state-of-the-art approaches
in co-execution, and related work in heterogeneous matrix multiplication. In Section~\ref{sec:poas}, we present POAS, our general framework for allowing co-execution
in heterogeneous environments. We provide an application example of POAS to a real-world application like the matrix multiplication in Section~\ref{sec:poasgemm}.
A performance evaluation of the matrix multiplication implementation of POAS is shown in Section~\ref{sec:evaluation}. Finally, Section~\ref{sec:conclusions} concludes
the paper and gives some hints for future work.

\section{Background and related work} \label{sec:background}
\subsection{Scheduling and co-execution}
Task scheduling is a research field that has been intensively studied over time.
Another related topic with scheduling, co-execution, has emerged in recent years motivated by the rapid hardware evolution.
In task scheduling, the objective is to distribute a set of tasks among different devices (task parallelism). 
This idea works well if assuming that all the devices within a system have good capabilities to perform any of these tasks.
In a world dominated by accelerators, this is not the case, as each of them is designed to work in restricted domains.
Henceforth, co-execution appears to be more suitable, as it exploits parallelism using multiple devices (accelerators) but only for a given task (data parallelism).
Ideas from both approaches often overlap, so it is worth studying works from both areas.

Performance prediction is a commonly used approach in scheduling, and it has been extensively studied in the last decades~\cite{surveyperfpred}.
This technique consists in predicting the application performance for a given set of hardware resources, allowing to estimate the execution time of an application.
In this context, we can differentiate between analytical (mathematical models) and non-analytical methods, which often rely on machine learning techniques.
However, using standalone analytical or non-analytical models can barely be used to predict the performance, so they are typically coupled with some characterization. For example, executing total or partially the application to help deduce its behavior, analyzing the source code, or carrying out some profiling. 
This characterization may provide information about the hardware and/or software to be measured.
In this sense, the roofline model~\cite{roofline} may also be useful to understand the performance behavior.
With the heterogeneous computing growth in last years, it has also been proposed for GPUs~\cite{rooflinegpu} but also SoCs with different accelerators~\cite{rooflinesoc}.

Task scheduling techniques have been proposed for OpenCL kernels in~\cite{smartopencl}, where authors use both static code features as well as runtime ones to predict the speedup of applications in CPU or GPU.
Also in OpenCL, non-analytical methods like decision-based trees are used in~\cite{mergeorseparate} to schedule OpenCL kernels on CPU/GPU platforms.
Co-execution opportunities are studied in~\cite{integratedcpugpu} on integrated CPU/GPU architectures.
They also study how to determine which compute elements are suitable or not for a given task (in other words, when co-execution is beneficial or not).
List scheduling has been applied in both static~\cite{task2021} and dynamic runtime scenarios, where new workloads arrive over time~\cite{schedsoc}.
Profiling and machine learning is combined in~\cite{geng2022} to provide scheduling in heterogeneous environments.
Integer linear programming (ILP) and linear regression are combined with stream graphs in~\cite{nguyen2016} to efficiently distribute workloads on multi-GPU platforms.
Performance modeling has been widely applied in many works~\cite{balanceperfmodel,corecomb,prem,heprem}.
In a DynamIQ heterogeneous multi-core environment, a performance model to estimate the efficient distribution of critical sections is designed~\cite{corecomb}.
Task scheduling has been often applied to CPU/GPU environments, but there are also other approaches for more heterogeneous environments, like CPU/FPGA~\cite{fpgasched}.
In~\cite{yesil2022}, authors propose a scheduling strategy for distributed accelerator-rich environments centered in real-time applications.
The predictable execution model (PREM)~\cite{prem} was proposed to enable time prediction on non-predictable hardware.
The approach separates programs into memory and compute phases, which can be independently scheduled.
It was proposed for CPU only, but a recent work extended it for CPU/GPU architectures~\cite{heprem}.
Many of these works focus primarily on minimizing the execution time. 
However, given the heterogeneous nature of today's computing systems, other studies consider both execution time as well as energy consumption in their scheduling decissions~\cite{peng2022}.

\subsection{General methods for co-execution} \label{subsec:coexe}
Several works have focused on designing a general method to provide co-execution of any data-parallel workload.
They are usually targeted to specific frameworks or languages that enable single-source coding on heterogeneous platforms.
One of such languages that are lately gaining influence is oneAPI~\cite{oneapi}.
However, oneAPI does not officially provide a mechanism for co-execution.
In a recent research~\cite{coexeoneapi}, authors proposed a new co-execution runtime in oneAPI based on load balancing algorithms. 
Another relevant framework in this context is OpenCL, which also was coupled with a co-execution engine in~\cite{enginecl}.
Lastly, in~\cite{ompsssched}, the authors extend the OmpSs framework to allow co-execution of OpenCL kernels.
General methods like the previously mentioned are particularly challenging as they need to efficiently schedule unseen kernels that may have potentially unbounded different behaviors.

Mentioned works have focused on co-executing generic kernels without domain-specific knowledge of the application being executed. 
This goal is quite hard to accomplish, as the framework must dynamically adapt the work depending on the kernel and hardware properties. 
It is more suitable for device-agnostic languages like OpenCL but, due to its high abstraction of the running code, it is unable to apply domain-specific optimizations on the work distribution. 
For these approaches, one of the main ways to know how to schedule the application is the source code.
However, the behavior of each application might be radically different, and predicting its behavior beforehand is not always feasible.
Instead of using performance prediction, other approaches like queue or list-based scheduling seem more appropriate in this context.
Yet, these techniques can easily suffer from load imbalance and often provide sub-optimal scheduling solutions.

\subsection{Heterogeneous matrix multiplication}
General matrix multiplication is a topic that has been deeply studied over time, mainly due to its high relevance in many computer science applications.
In this sense, one of the main concerns is matrix multiplication performance.
Recent works have studied the performance of matrix multiplication in heterogeneous environments~\cite{compareperfgemm}.
Furthermore, several papers have considered the use of different hardware devices to compute matrix multiplications to exploit heterogeneous systems.
One of the first studies~\cite{beaumont2001} already approached the problem from an analytical point of view.
The authors analyzed the computational power of each processor in the heterogeneous system and later expressed the workload distribution as an optimization problem.
As the concept of heterogeneity has evolved, that work was targeted to distributed systems using MPI, which imposed different issues to be solved.
In~\cite{camarahierarchies}, the authors designed a hierarchical approach to be able to distribute parts of the matrix multiplication to different devices.
Considering multiple accelerators and a range of $n$ columns to be assigned to each accelerator, the search space becomes too big.
Therefore, they proposed a hierarchical way of considering all the possibilities, significantly reducing the search space.
A new algorithm based on Strassen's method was presented in~\cite{hpmax} for heterogeneous environments.
To schedule the work between accelerators, a queue-based system was used, which gives blocks of the matrices to be computed whenever a device is free.
Matrix multiplication workload distribution has also been studied in the context of energy efficiency~\cite{catalan2015}, where authors proposed an approach for ARM big.LITTLE processors.
One of the centric ideas was to study the performance ratio between the big and the little cores in the SoC and use such criteria to perform the scheduling.

\subsection{Accelerators and tensor cores}
Accelerators are hardware devices that execute a given workload in less time and/or with much less energy than conventional CPUs~\cite{acmacc}.
Nowadays, GPUs are the mainstream, easily accessible accelerators for the masses.
While they accelerate many relevant workloads (like machine learning)~\cite{evoGPU}, they are still generic enough to be used for many domains.
However, there is a trade-off between efficiency and generality, so GPUs are usually less efficient compared to more specific accelerators.
FPGAs can be adapted to different specific domains thanks to their re-programmable hardware.
Lastly, application-specific integrated circuits (ASICs) are designed and built for specific applications, so they achieve the highest levels of performance and efficiency.
Popular domains have already plenty of accelerators designed for them, as is the case of machine learning.
The well-known tensor processing unit (TPU)~\cite{tpu} accelerates both inference and training workloads.
Many other accelerators exist in this field for this task, but also other even more domain-specific ones, like neural radiance field~\cite{icarus}.
In the area of matrix multiplication, accelerators supporting dense and sparse products~\cite{gemmaccdense}, as well as sparse only matrix multiplication~\cite{gemmacc} exists.

A trend in computer architecture to make accelerators more accessible to broader domains is to incorporate domain-specific cores in general-purpose processors~\cite{acmacc}.
This allows using the general processor for generic tasks while offloading domain-specific ones to specialized cores.
The tensor cores~\cite{voltauarch} included in Volta, Turing and Ampere microarchitectures are a good example of this idea.
Tensor cores debuted in Volta microarchitecture~\cite{dissectingVolta}, which have evolved and improved its performance with newer generations.
This domain-specific hardware is primarily designed to enhance the performance of deep learning applications, but can also be used to enhance pure matrix multiplication workloads.
In Volta, tensor cores implement a 4x4x4 FP16 matrix multiply and accumulate instruction, HMMA (half precision matrix multiplication and accumulate)~\cite{dissectingVolta}.
The Turing tensor cores adds support for \texttt{int8}, \texttt{int4} and \texttt{int1} data types~\cite{dissectingTuring} through a new IMMA instruction.
In Ampere microarchitecture, the matrix multiplication size changes from 4x4x4 to 8x4x8, doubling its FP16 throughput~\cite{evoGPU}. 
It also adds new instructions for sparse matrix multiplication, which in turn doubles the throughput of dense matrix multiplications.
Tensor cores boost specific applications' performance in an unprecedented way, providing a 4x boost in peak performance compared to CUDA cores, and 8x for the case of sparse matrices~\cite{evoGPU}.

\section{Predict, Optimize, Adapt and Schedule (POAS)} \label{sec:poas}
\begin{figure*}[ht]
  \centering
  \includegraphics[width=\textwidth]{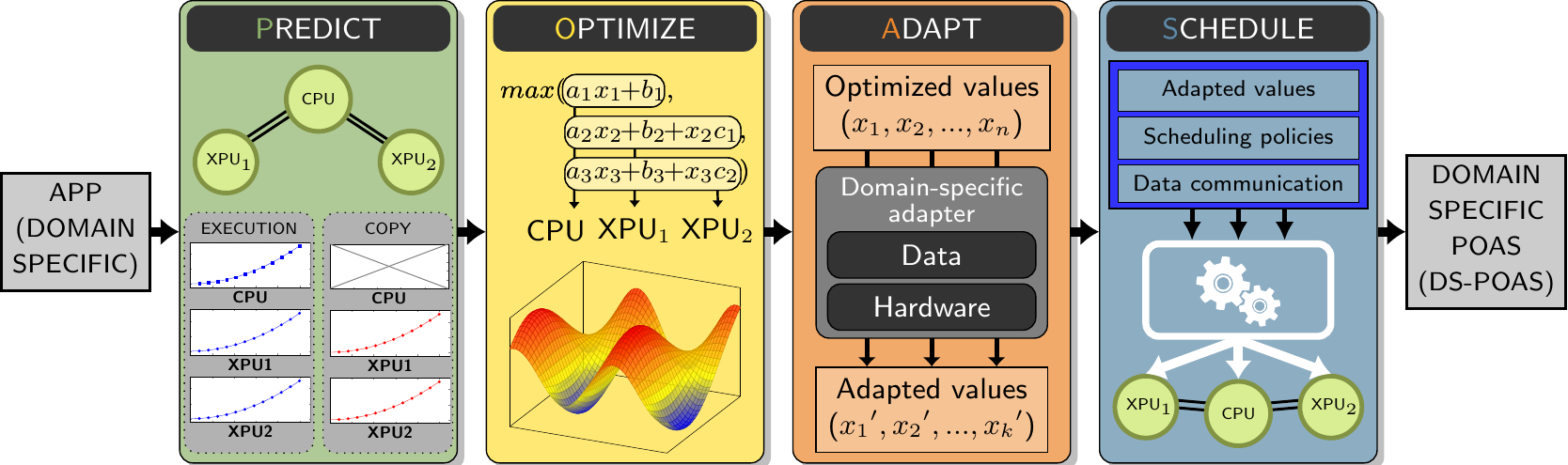}
  \caption{General overview of the POAS framework. POAS (Predict, Optimize, Adapt and Schedule) takes a generic application and adapts it into a DS-POAS (domain-specific POAS), the same application but ready for co-execution, allowing ALP in heterogeneous environments.} \label{fig:poas}
\end{figure*}

POAS (Predict, Optimize, Adapt and Schedule) is a generic framework that can be adapted to any application domain.
It is worth noting that POAS is not a framework that automatically schedules any generic application, but a generic model that allows defining domain-specific solutions to schedule any application.
Therefore, POAS must first be adapted to the specific application domain.
The result of adapting POAS to a specific application is a DS-POAS (domain-specific POAS).
A DS-POAS is a model that takes the original application and adapts it to provide co-execution.
On the one hand, the POAS model requires an additional step to be able to work in a given scenario (manually designing the specific DS-POAS), which requires extra work compared to generic approaches.
On the other hand, as the resultant model has domain-specific information, it can provide more precise scheduling solutions.
With a present and a future dominated by domain-specific hardware, a domain-specific model for co-execution like POAS can be crucial to allow accelerator-level parallelism in modern hardware.
To the best of our knowledge, this is the first work that proposes a generic scheduling framework for heterogeneous scenarios that can be adapted to any domain, and the first approach that divides the scheduling problem into four differentiated parts.

A general view of POAS is depicted in Figure~\ref{fig:poas}.
The POAS model is divided into four phases (Predict, Optimize, Adapt and Schedule), which must be performed in order.
In this sense, the output of each phase is the input of the next one, as Figure~\ref{fig:poas} shows.
The model can be adapted to achieve different goals in ALP environments, like minimizing the execution time or the energy used.
Lastly, the framework was designed to work in scenarios where there is a significant amount of work so that ALP can be exploited to achieve some profit.
It does not detect whether an application is suitable for co-execution or not, which is something that we leave for future work.

\subsection{Predict}
In the predict phase, a performance predictor is designed.
The goal of the performance prediction is to give a precise estimation of the execution time (or energy needed, if the model wants to be optimized for energy efficiency) of the studied application.
This prediction is both software and hardware-dependant, so the prediction must consider both application and hardware characteristics.
As part of the prediction phase, a profiling of the hardware platform is also necessary.
With profiling, the predictor can seamlessly adapt to the hardware and software, providing high precision.
It is essential to carefully study the behavior of the hardware executing the application, as sometimes it provides different performance results depending on data sizes, alignment, and other factors.
If possible, profiling should be performed in the optimal conditions of the harwdare, as real workloads would be performed in the same conditions.
In either case, it is crucial is to ensure that profiling and real workloads are executed under the same conditions because if this is not the case, the prediction would be suboptimal.
Besides, it is worth noting that for POAS it is not mandatory to have the source code, which is a limitation that exists in many language-centered models.

The POAS framework is modular, so any performance prediction method can be chosen in this phase.
There are many performance prediction approaches, and depending on the domain, one or another might be more suitable in each case.
Furthermore, the performance model must predict not only the execution time but also the time spent in memory transfers between the CPU and the accelerators.
The only requirement for the prediction is to provide a mathematical function that, given the input size, predicts the execution time of the application for a variety of hardware devices.
While the resultant function has no restriction regarding its complexity, it is desirable to have a linear or quadratic function, as discussed in Section~\ref{sec:poas2}.
The prediction accuracy is crucial to achieve competitive performance; if the prediction fails to precisely reproduce the experimental results, the scheduling will be poor.

\subsection{Optimize} \label{sec:poas2}
The optimization phase takes the prediction model generated in the previous step as input.
This phase has two objectives: to define a formulation of the behavior of the application and how to optimize it.
The formulation is enunciated as a constraint satisfaction problem (CSP), which can be formulated to minimize different objectives (like execution time or energy consumption).
In many cases, however, the problem can be further specialized into a constrained-optimization problem (COP), which is a generalization of the CSP.
It is crucial that the mathematical formulation models all the details of how the application works in the real world (i.e., when the compute and communication phase occurs and how).
Regarding methods for optimizing the model, linear or quadratic programming can be used, providing the optimal solution in very little time.
However, these methods can only be used if the function that models the behavior of the application is linear or quadratic.
Because there might be cases where the performance model is too complex to be represented in these terms (e.g., the function is cubic), the problem should be formulated as a CSP. In this case, alternative methods like backtracking or local search could be used to optimize the performance model.
Anyhow, the result of this phase is a set of optimized values, which typically represent the input size of each device, such that the desired objective function is optimized.

\subsection{Adapt}
Depending on the application, the variables that come from the optimized model designed in the previous phase might need some transformations to be used by the scheduler.
Therefore, an intermediate phase called adapt is needed to make the scheduler work correctly.
We differentiate between two types of adjustments: data and hardware adjustments.

\subsubsection{Data adjustments}
The output of the optimized model may contain different information than the one needed to determine how to schedule the application.
For example, it might be necessary to abstract the data that the mathematical models work with, adding, removing variables, or altering in any other form the data used.
Adding abstraction or changing the variables of the problem formulation might be beneficial in some cases.
It might be useful to reduce the complexity of the problem (from cubic to linear models) to improve the prediction performance or to solve application-specific problems.
If such alteration occurs, the output of the optimization phase cannot be used to decide how to do the scheduling.
In these cases, the adapt phase must adjust the values given by the optimization phase to some values that can actually be used in the scheduler.
Since data adjustments depend on the application, this procedure is essentially application-dependent.

\subsubsection{Hardware adjustments}
Generally speaking, hardware is very sensitive to data sizes and other factors, so optimal performance is only reached in given circumstances.
These circumstances heavily vary depending on the hardware platform.
As we mentioned before, the preferred choice for efficient use of hardware is to profile the platform assuming optimal input values.
If the profiling was not performed assuming optimal performance, the hardware adjustments must also match the same conditions.
In either situation, the solution given by the optimized model may not conform to the conditions in which profiling was done.
Therefore, fine-tuning must be performed to adapt the output of the model to input sizes that efficiently use the hardware.
On the one hand, it is necessary because we are interested in maximizing throughput or minimizing energy consumption, which is only achieved by fulfilling this goal.
On the other hand, it is necessary because if profiling was performed assuming optimal values but real workloads are not executed in the same conditions, prediction accuracy would be poor.

\subsection{Scheduler}
Within the POAS model, the scheduler can work in two different ways: static and dynamic.
Other scheduling policies, as well as modifications to the presented ones, are left to future work.
Furthermore, the scheduler must also consider how to schedule the communications between the CPU and the accelerators, which might have a significant impact on performance.

\subsubsection{Static scheduling}
The static scheduler uses the performance model and optimizes the problem formulation once to get the optimal inputs for each device.
It is the simplest mode as it does not change over the execution of the program.
This mode works well when the application requirements do not change over time, and when the performance prediction can model precisely the behavior of the hardware.
If one of these requirements do not meet, the static scheduling would provide inaccurate predictions of the execution time of the application, leading to suboptimal scheduling where the hardware utilization may decrease significantly.

\subsubsection{Dynamic scheduling}
To overcome the aforementioned problems, a dynamic scheduler can be employed.
In the dynamic scheduler, the performance prediction model is used to optimize the function and obtain the optimal values, just like the static scheduling.
But, unlike static scheduling, any of the other three phases (Predict, Optimize or Adapt) may be changed over time, thus modifying the performance prediction, the optimization function, or the application adapter.
If an application performance varies over time (e.g., hardware can be added or removed dynamically, or because performance heavily depends on external factors), the performance prediction could be altered during execution.
To do so, one approach is to be constantly measuring the execution time of the application and adapting the performance model over certain periods, which granularity can be adjusted as needed (e.g., every second, or every program iteration).
Furthermore, it is also possible that the problem changes during execution, which may need the constraint problem to be reformulated.

\subsubsection{Data communication scheme}
One performance crucial aspect of the work distribution is the effective use of the memory bus.
In ALP environments (like SoCs), accelerators are typically connected to a shared bus, where all of them can communicate with the CPU.
Because of that, optimizing applications for exploiting ALP is challenging since the bus (thus, the throughput) must be shared between all the accelerators.

As a first approach, we propose a scheduler based on priority scheduling.
The idea is to assign a priority to each device connected to the shared bus.
Then, data is copied to/from the CPU in the order dictated by the priority ordering.
There are many approaches to designing this scheme with different goals, like minimizing the idle time of accelerators.
We leave for future work to further investigate more efficient approaches.

\section{Applying POAS to GEMM (hgemms)} \label{sec:poasgemm}
This section details how POAS can be adapted to a specific application and provides concrete examples for all of the four phases in POAS.
The result of this section is a DS-POAS specialized for matrix multiplication, which we will be referring to as \texttt{hgemms} (heterogeneous GEMM scheduler).
Due to the modular nature of POAS, the presented DS-POAS is one of the multiple possible DS-POASs specialized for this application.
The goal of this section is not to show the optimal approach to matrix multiplication, but to present an example of POAS applied to a real-world application.

We designed \texttt{hgemms} for minimizing the execution time and targeted CPUs, GPUs and tensor cores (from now on, XPUs).
The implementation relies on optimized libraries to perform the matrix multiplications: MKL (in Intel CPUs), BLIS (in AMD CPUs) and cuBLAS (for both CUDA and tensor cores).

\subsection{Predict} \label{subsec:gemmpredict}
\subsubsection{Linear regression}
To design the performance predictor for GEMM, we used a regression analysis approach.
It is well known that the general algorithm has a complexity of $O(n^3)$.
But to use linear regression, we must find a way to represent the time with linear growth.
We acomplish this by modeling the execution time with the number of operations (from now on, \textit{ops}), such that $ops = m*n*k$, instead of the input size ($n^3$).
In other words, the execution time growths with a cubic complexity if we consider the input size, but grows linearly considering the number of operations.

While this linear function can generally predict the performance of GEMM, there exist certain hardware peculiarities which might cause the prediction to fail.
For example, the XPU will provide radically different results depending on the input size of the matrix, as the tensor cores can only be optimally used when the input meets some criteria.
To eliminate ambiguity, the performance predictor always assumes optimal performance.
In other words, if a device only reaches optimal performance under certain conditions, the predictor will always predict the execution time under such circumstances.
Therefore, one additional task of \texttt{hgemms} in the adapt phase is to ensure that the real workloads can actually be computed in the same way as the predictor was trained for.
We further contemplate these details in Section~\ref{subsec:gemmadapt}.
In addition to the compute times, \texttt{hgemms} also predicts copy times between CPU and GPU.

\subsubsection{Profiling}
We perform a profiling of the hardware platform, which is done only once at installation time and takes less than five minutes to complete.
The profiling phase measures the compute power of all the hardware devices available in the system and the memory bandwidth between CPU and GPUs.
Then, the results are stored in a text file that is read when real matrix multiplication workloads arrive.
\begin{itemize}
\item Computing power profiling: The program runs a set of squared matrix multiplications (using appropiate libraries like MKL, BLIS or cuBLAS).
The sizes of the squared matrices are variable and adjustable depending on the device (see Section~\ref{sec:evprofiling} for more details).
When all the experiments have finished, linear regression is performed to obtain the linear function that models the execution time of the device.
\item Memory bandwidth profiling: The program runs a microbenchmark that measures the bandwidth between the CPU and each of the GPUs.
\end{itemize}

One limitation of this approach is the fact that the profiling only collects data of squared matrices.
Depending on the hardware platform and the library used, a matrix multiplication with the same number of operations but different shape might need different amounts of time to complete.
However, profiling the performance of any matrix size is not feasible, as it would require large amounts of time to complete.
Our solution is to consider that any matrix multiplication can be decomposed into sub-matrices products, whose number of operations is equal to the sum of the operations of the sub-matrices products.
The key idea is that the sub-matrices can be decomposed to be squared while maintaining the same number of operations.
Using this approach, we can also predict the performance of non-squared matrices precisely.

\subsection{Optimize} \label{sec:gemmsolve}
In the optimization phase, we formulated a constraint satisfaction problem that minimizes the execution time.
Therefore, the goal of the solver is to find a distribution of \textit{ops} among the hardware devices such that the total execution time is minimal.

\subsubsection{Problem formulation}
We express the execution and copy times as a mixed-integer linear programming (MILP) problem.
We define $c_x$ as the independent variables, which represents the number of operations (\textit{ops}) to be computed by device $x$.
The goal of the solver is to minimize the objective function (which models the total execution time of the GEMM in $n$ devices):

\begin{align}
max(t_{c_1}+t_{y_1}, t_{c_2}+t_{y_2}, ...\ , t_{c_n}+t_{y_n}) \label{eq7}
\end{align}

where:
\begin{itemize}
\item $n$ is the number of devices in the system.
\item $t_{c_x}$ is a linear function in the form $ac_x+b$ that models the execution time of the device $x$ when it computes $c_x$ operations.
\item $t_{y_x}$ is a linear function that models the copy time of the device $x$ when it computes $c_x$ operations (if $x$ is a CPU, then $t_{y_x}=0$).
\end{itemize}

with constraints:

\begin{align}
c_1,c_2,...\ ,c_n &\geq 0 \\
\sum_{i=0}^{n}c_i &= N
\end{align}

where $N$ is the total number of operations to be computed (i.e., $m*n*k$), and with the copy time functions defined as:

\begin{align}
y_x = \frac{dt_x*c_x\left(\frac{1}{k}+\frac{1}{n}\right)+kn}{bw_x} \label{eq4}
\end{align}

with:

\begin{itemize}
\item $dt_x$ being the data type size in bytes of device $x$.
\item $k,n$ being the sizes of the matrix (constants).
\item $bw_x$ being the memory bandwith between CPU and the device $x$, measured in bytes per second.
\end{itemize}

Equation~\ref{eq4} gives the time to copy $A$, $B$ and $C$ matrices, assuming that the communications happens in a bus exclusively used by device $x$.
This is true when only one device is connected to the bus, but is not realistic in a shared bus (e.g., in a SoC).
If memory copies of different devices are serialized, the function must take into account the time to copy the data of previous devices of $A$, $B$ or $C$ matrices.
We modified the equation in our formulation as our target platform includes more than one accelerator connected to the same bus.
Besides, the memory prediction model is simplified, as it only considers the memory bandwidth but not the latency.
As mentioned, POAS should only be applied to applications where there is significant compute work.
Thus, since we apply \texttt{hgemms} to scenarios that involve large data transfers, the latency is unnoticeable.

We implemented the MILP problem using CPLEX 12.10~\cite{cplex}.
The CPLEX solver is embedded in the program using the CPLEX API, and the MILP formulation is dynamically defined depending on the devices being used.
When the model has been optimized, the output variables of the MILP solver are $c_1, c_2, ...\, c_n$, which represent the number of operations that each device has to compute.

\subsection{Adapt} \label{subsec:gemmadapt}
In the adapt phase, the optimized values given by the MILP solver must be adapted to be used by the scheduler.
For this task, we designed an algorithm called \texttt{ops\_to\_mnk} that works on both data and hardware adjustments.

\subsubsection{Data adjustments} \label{sec:findmnk}
To properly schedule parts of the matrices to be computed on each device, we need concrete matrix sizes (i.e., $m$, $n$ and $k$).
Having the output vales instead of the $m$, $n$ and $k$ dimensions might appear a disadvantage, because an algorithm is needed to do the mapping but it might become an advantage, as it provides flexibility.
Regarding data adjustments, the \texttt{ops\_to\_mnk} algorithm must accomplish two tasks:

\begin{enumerate}
\item Find $m,n$ and $k$ such that the number of operations matches the operations given by the MILP solver. 
This gives the $m$, $n$ and $k$ dimensions for each device.
\item Express the global matrix product as a list of squared sub-matrices products (in a best-effort manner).
This divides the $m$, $n$ and $k$ dimensions for each device into sub-matrices for precise performance prediction.
\end{enumerate}

For the first task, we start setting $n$ and $k$ to their original values.
Partitioning a matrix with a different value of $n$ would provide partial results in the output $C$ matrix, so we fix $n$ for conviniency.
Setting $k$ to the original value makes the \texttt{ops\_to\_mnk} algorithm easier since just the rows must be distributed.
Then, to map \textit{ops} to $mnk$, only $m$ has to be determined, which is computed as $m = \frac{ops}{n*k}$.

For the second task, the algorithm must ensure that resultant matrices are as squared as possible (best-effort).
Having squared matrices is the optimal scenario, as we would be performing the matrix multiplications in the same way as the prediction did.
But this can only be accomplished if the input size is divisible by the sub-matrix sizes, which is not always possible.
However, near-squared matrices (e.g., $m=1.1k$) can also be predicted with very high precision.
Let us denote with an apostrophe the dimensions of the submatrix (e.g., $k'$) and without it, the dimensions of the original matrix (e.g., $k$).
The algorithm tries to make $m'$ and $k'$ as similar as possible, while keeping $n'$ equal to $n$.
Our algorithm always ensures that the number of horizontal dimensions in $A$ fits perfectly (i.e., $k\ \%\ k' = 0$).
Without such restriction, ``gaps" may appear in the last column of $A$.
Therefore, the search space in $k'$ is restricted to the divisors of $k$, which happens to be big enough when the input matrix is also big.
For determining $m'$ size, the algorithm iterates over all the possibilities, analyzing how ``squared'' would be the resultant matrices using a simple heuristic.
For a given list of squared matrices with $\{m_1',m_2',...\ ,m_n'\}$ and $\{k_1',k_2',...\ ,k_n'\}$, the squareness (\textit{sq}) is computed as:

\begin{align}
sq = \sum_{i=0}^{N}\left(\frac{min(m_i',k_i')}{max(m_i',k_i')}*m_i'k_i'n\right)
\end{align}

This value represents how squared is the global set of sub-matrices.
Thus, to find the best sub-matrix distribution, the algorithm chooses the one that maximizes the value of the heuristic.

\subsubsection{Hardware adjustments}
The \texttt{ops\_to\_mnk} algorithm asserts that the matrix sizes satisfy the requirements imposed by the hardware to achieve optimal performance.
In our case study, we consider CPUs, GPUs and tensor cores, so the \texttt{ops\_to\_mnk} algorithm must meet two requirements:

\begin{itemize}
\item Tensor Cores: To reach optimal performance, the input sizes must meet the following conditions: $m\ \% \ 8 == 0$ and $k\ \% \ 8 == 0$~\footnote{See \url{https://docs.nvidia.com/cuda/cublas/index.html\#tensorop-restrictions} for more details}.
To do so, the algorithm reduces the input size until it meets the desired requirements.
In the end, this means that the tensor cores get a fewer operations than the MILP solver specified, but this is barely unnoticeable since the size reduction is tiny compared to the global size.
\item CPU cores: When profiling the CPU, inputs are designed to fit into cache memory. 
Therefore, when a real workload arrives, the algorithm must ensure that the generated submatrices also fit into cache. 
\end{itemize}

\subsection{Scheduler}
\begin{figure}[t]
\includegraphics[width=\linewidth]{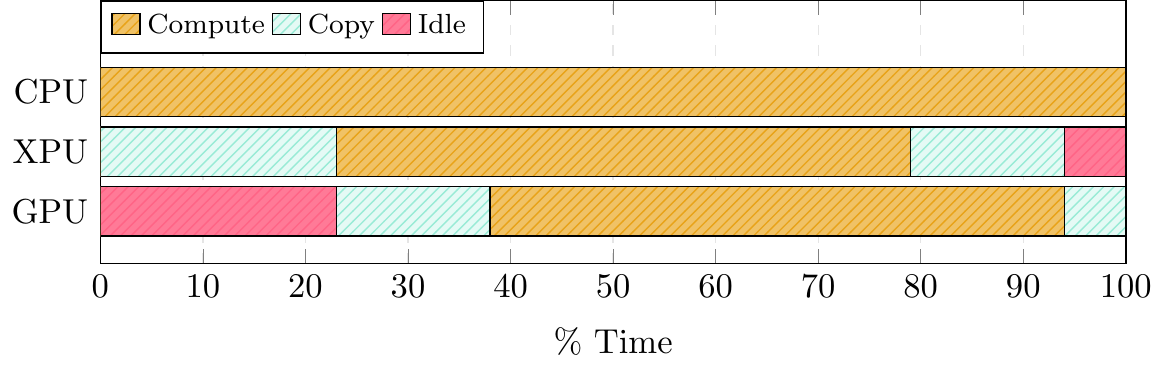}
\caption{Proposed scheduling communication scheme in a shared bus with CPU+GPU+XPU} \label{fig:comm}
\end{figure}

For the scheduler, we used a static scheduling, as we found that gives excellent results for our case study.
The scheduler receives the matrix sizes for each device and does not change them over time.
We explore some of the possible issues of this approach in Section~\ref{sec:performance}.

Regarding the use of the shared PCIe bus, we used a priority scheduling approach.
When the program reads the configuration file, it assigns a priority for each device: the faster the device, the higher priority.
Then, $A$ and $B$ matrices are copied in the order established by the priority.
Thus, lower priority accelerators remain idle while the higher priority devices are copying the data.
After the computation, the first device (meaning the faster one) copies $C$ to the host, and the same order is used to copy the remaining parts of $C$.
In this case, the higher idle times are experienced from high-priority devices, which have to wait for the rest of the devices to complete.
Figure~\ref{fig:comm} shows the proposed communication scheme.

\subsection{Implementation details}
When tensor cores are used, the output of the matrix multiplication comes in half-precision (FP16), while the CPU (which uses the MKL or BLIS library) can only perform the product in FP32.
Therefore, when the results are collected in the CPU, the product has mixed-precision results. 
In this work, we do not consider how to deal with this problem as it is out of the scope of the paper, but it is worth mentioning that related work in this field has shown promising results~\cite{hmmafp32}.

In our implementation, we copy the data between CPU and GPU synchronously.
This simple approach could be improved using CUDA streams and overlapping the computation with memory copies.
In either case, the performance predictor can be adapted to predict the memory copies with or without overlap.
Therefore, for our study, it is not particularly relevant whether the implementation copies the data with or without overlap.

\section{Evaluation} \label{sec:evaluation}
This section evaluates our proposal applied to GEMM in two ways: prediction accuracy, in Section~\ref{sec:accuracy} and performance, in Section~\ref{sec:performance}.
\subsection{Test bed}
\subsubsection{Hardware and software configuration}
\begin{table}[t]
\centering
\resizebox{\linewidth}{!}{%
\begin{tabular}{lllll}
\toprule
                 & \multicolumn{2}{c}{CPUs}          & \multicolumn{2}{c}{GPUs / XPUs} \\ \cmidrule(l){2-3} \cmidrule(l){4-5}
Model            & \begin{tabular}[c]{@{}c@{}}Intel Xeon\\ E5-2603 v3\end{tabular} & \begin{tabular}[c]{@{}c@{}}AMD\\EPYC 7413\end{tabular} & \begin{tabular}[c]{@{}c@{}}NVIDIA\\ RTX 2080 Ti\end{tabular} & \begin{tabular}[c]{@{}c@{}}NVIDIA\\ RTX 3090\end{tabular}       \\ \midrule
Architecture     & Haswell         & Zen 3         & Turing        & Ampere         \\
Technology       & 22nm            & 7nm           & 12nm          & 8nm            \\
CPU cores        & 6               & 24            & -             & -              \\
CUDA cores       & -               & -             & 4352          & 10496          \\
Tensor cores     & -               & -             & 544           & 328            \\
Max. Frequency   & 1.6 GHz         & 3.6 GHz       & 1.5 GHz       & 1.6 GHz        \\
TFLOP/s (FP32)   & 0.307           & 2.76          & 13.45         & 35.58          \\
TFLOP/s (FP16)   & -               & -             & 107.5         & 284.65         \\
LLC              & 15MB            & 128MB         & 6MB           & 6MB            \\
Memory size      & 64GB            & 512GB         & 11GB          & 24GB           \\
\bottomrule
\end{tabular}
}
\caption{Hardware specifications for the testbed environment} \label{tab:testbed}
\end{table}

\begin{table}[t]
\begin{tabular}{llll}
\toprule
      & CPU       & GPU         & XPU         \\ \midrule
mach1 & Xeon v3   & RTX 2080 Ti & RTX 2080 Ti \\
mach2 & AMD EPYC  & RTX 3090    & RTX 2080 Ti \\
\bottomrule
\end{tabular}
\caption{Hardware configuration for the two testbed environments} \label{tab:config}
\end{table}
The evaluation platform is represented by mach1 and mach2, two HPC servers with a CPU+GPU+XPU configuration.
During this evaluation, we refer to an XPU as the GPU that uses the tensor cores to perform the matrix multiplication, whereas GPU uses traditional CUDA cores.
The hardware configuration of both machines is summarized in Table~\ref{tab:config}, and the specifications for each device are detailed in Table~\ref{tab:testbed}.

Both systems ran Centos 8.2 (4.18.0-193 kernel in mach1 and 4.18.0-348 in mach2).
The source code of \texttt{hgemms} was built using g++ 8.4.1.
For running the GEMM in the Intel CPU, we used Intel MKL version 2020.0.2.
For the AMD CPU, we used AMD AOCL BLIS 3.1.
For the GPUs and XPUs, mach1 ran the NVIDIA driver 450.51 and mach2 the version 510.47, while the NVIDIA cuBLAS version employed in mach1 was the 11.2.0, and the 11.8.1 in mach2.
Regarding the communication between CPU and GPUs, the RTX 2080Ti's in mach1 are connected to a PCIe 3.0 x16 bus, which peak memory bandwidth is 15.75 GB/s.
In mach2, both cards are connected to a PCIe 4.0 x16 bus, providing a peak memory bandwidth of 31.75 GB/s.
Since the RTX 2080Ti supports up to PCIe 3.0, the card in mach2 works in 3.0 mode, even though it is connected to a 4.0 slot.
For the experiments, we reserved one physical CPU core for managing the GPU and XPU. 
Thus, in our experiments, mach1 worked with 5 physical cores and mach2 with 23.

\subsubsection{Matrix sizes}
To evaluate our approach, we conceived six different matrix sizes, which are shown in Table~\ref{tab:inputs}, sorted in descending order by the number of operations (TOps).
The first input size evaluates the performance for a relatively small squared matrix.
A larger non-square matrix is evaluated with the second input.
We are also interested in evaluating very skinny matrices like input 3, where the $m$ dimension is much larger than the others.
We explore the same idea for the $n$ dimension in input 4, and $k$ in input 5.
Lastly, input 6 is the biggest matrix multiplication in the list.

\begin{table}[t]
\begin{tabular}{rrrrr}
\toprule
   & $m$  & $n$ & $k$ & TOps \\ \midrule
i1 & 30K  & 30K & 30K & 27.0 \\
i2 & 60K  & 20K & 35K & 42.0 \\
i3 & 130K & 20K & 20K & 52.0 \\
i4 & 40K  & 80K & 20K & 64.0 \\
i5 & 40K  & 30K & 60K & 72.0 \\
i6 & 56K  & 40K & 40K & 89.6 \\
\bottomrule
\end{tabular}
\caption{Matrix sizes used for the evaluation} \label{tab:inputs}
\end{table}

We repeated the computations for each input 50 times, therefore executing 50 matrix multiplication over the accumulated data.
We used these inputs for Sections~\ref{sec:accuracy} and ~\ref{sec:performance}.
Each input was run 3 times, and the values shown in both sections are the average over these 3 independent runs. 

\subsubsection{Profiling and hgemms configuration} \label{sec:evprofiling}
The profiling phase performed 30 squared matrix products with matrix sizes ranging between 1000 and 2000 for the CPU and between 3000 and 6000 for GPU/XPU.
Each matrix product is measured 5 times, and the average over these 5 independent runs is used to perform the linear regression.
For the generation of the list of squared sub-matrices, they were restricted to be of a size such that the number of operations were between the same number of operations that were performed during profiling.
In other words, in CPU the sub-matrices were restricted to 1000\ x\ 1000\ x\ 1000 ($10^9$) and 2000\ x\ 2000\ x\ 2000 ($8 * 10^9$) operations, and in GPU between 3000\ x\ 3000\ x\ 3000 ($27 * 10^8$) and 6000\ x\ 6000\ x\ 6000 ($216 * 10^8$) operations.
Thus, sizes are computed on the fly depending on the size of $n$ in the original matrix.

\subsection{Prediction accuracy} \label{sec:accuracy}
To evaluate the performance predictor used in \texttt{hgemms}, we measured the prediction accuracy.
The execution and memory copy times are measured and compared to the predicted values.
We calculate the prediction error $e$ as an expression of the relative error: $e=100*\frac{v-v_{pred}}{v}$, where $v$ is the measured time in our experiments and $v_{pred}$ is the value given by the predictor.
We also calculated the root mean square error (RMSE), which gives a general perspective of the prediction robustness across different inputs.
The results are summarized in Table~\ref{tab:err}, and RMSE is shown in Table~\ref{tab:rmse}.

For GPU and XPU, we show the global prediction error in the first instance, followed by the computing and memory copy prediction error, respectively.
Overall, we observed that the prediction error is low (typically under 5\%).
This is a key factor to provide high-quality co-execution because otherwise, the load imbalance would be very high, leading to substantial performance degradation.
Except for a few cases, the memory prediction error is quite low, especially for mach2, which prediction is close to be perfect.
The major part of prediction error comes from compute prediction, which is much harder to estimate.
Some inputs were predicted with slightly higher prediction error ratios than the mean, especially the latest ones in the GPU and XPU in mach1.
In fact, these ``outliers'' are one of the facts that increase the RMSE of the whole evaluation.
We believe that this issue has a simple explanation that has to do with temperatures.
During benchmarking, we left all the devices' frequencies unlocked.
Depending on the platform, this means that the clock frequency can be downscaled significantly due to overheating. 
In other words, the measured frequency in the profiling phase may not match the frequency used in real workloads.
This is especially true for mach1 since it has significantly worse heat dissipation capabilities than mach2.
According to RMSE values in Table~\ref{tab:rmse}, \texttt{hgemms} achieves low RMSE values that confirm the great robustness of the predictor, despite using static scheduling.
However, a more sophisticated solution could employ a dynamic scheduler that considers the frequency in real-time of every device and dynamically balance the workload to further improve accuracy.

\begin{table}[t]
\resizebox{\linewidth}{!}{%
\begin{tabular}{rllllll}
\toprule
      & \multicolumn{3}{c}{mach1}       & \multicolumn{3}{c}{mach2} \\ \cmidrule(l){2-4} \cmidrule(l){5-7}
      & CPU    & GPU      & XPU     & CPU    & GPU     & XPU \\ \midrule
i1    & 4.5 & 1.6 (8.8,5.3)  & 0.7 (3.2,1.2) & 1.0 & 4.6 (9.1,0.0) & 4.7 (10.1,1.2) \\
i2    & 1.4 & 2.9 (5.1,0.6)  & 3.1 (6.1,0.1) & 0.5 & 1.6 (2.9,0.0) & 6.1 (11.6,1.0) \\
i3    & 3.1 & 0.7 (0.8,2.0)  & 3.3 (6.2,0.6) & 0.4 & 1.6 (3.3,0.0) & 7.4 (12.2,0.9) \\
i4    & 4.6 & 9.9 (5.3,14.4) & 5.3 (6.1,4.2) & 2.0 & 2.0 (3.9,0.1) & 4.6 (7.8,1.3)  \\
i5    & 2.4 & 6.9 (11.9,0.0) & 3.0 (6.6,0.1) & 1.3 & 5.8 (9.9,0.1) & 5.4 (10.7,1.1) \\
i6    & 0.8 & 6.5 (6.7,6.2)  & 3.4 (5.0,1.4) & 3.6 & 4.1 (6.8,0.0) & 6.7 (11.1,1.1) \\
\bottomrule
\end{tabular}
}
\caption{Prediction error for different input matrices in mach1 and mach2 (in \%)} \label{tab:err}
\end{table}

\begin{table}[t]
\begin{tabular}{rllllll}
\toprule
      & \multicolumn{3}{c}{mach1}       & \multicolumn{3}{c}{mach2} \\ \cmidrule(l){2-4} \cmidrule(l){5-7}
      & CPU    & GPU      & XPU     & CPU    & GPU     & XPU \\ \midrule
RMSE  & 2.42   &5.63      & 3.13    & 1.69	  & 2.85    & 4.42 \\
\bottomrule
\end{tabular}
\caption{Root mean square error (RMSE) for mach1 and mach2.} \label{tab:rmse}
\end{table}

\subsection{Performance} \label{sec:performance}

This section analyzes the performance improvements of \texttt{hgemms} respect to standalone execution (using only CPU, GPU or XPU).
Let us first analyze the workload distribution found by \texttt{hgemms}, which is shown in Table~\ref{tab:perc}.
We can rapidly conclude that the distribution changes vary very little from one input to another.
This is because all of the inputs are relatively big, and the computing power of accelerators oversees the memory copy overhead.
Depending on the matrix shape, the memory copy cost can vary significantly, which explains the different variations to find the optimal distribution between GPU and XPU.
Since the CPU is not affected by this issue, we can observe that its percentage of work distribution clearly decreases with increasing matrix sizes.
As the amount of computational work grows, the cost of memory copies is more profitable, making the compute work on the CPU less valuable.

In the end, this means that the speedup given by \texttt{hgemms} with respect to CPU and GPU is huge.
Compared to the XPU alone, \texttt{hgemms} provides a very good speedup in mach1, ranging from 14\% in the worst case up to 28\% in the best one.
We conclude that \texttt{hgemms} obtains decent speedup in mach1, given the big difference between the compute power of the XPU and the GPU and CPU.
A different situation is mach2, where \texttt{hgemms} reaches a 45\% of speedup in the best case.
In the end, performance gains are constrained by the relative computing power of the non-XPU devices compared with the XPU and memory throughput.
In mach1, the GPU contributes a good amount of computing power of the CPU+GPU+XPU aggregate, while the CPU gives almost no speedup.
In contrast, the CPU and especially the GPU in mach2 do contribute a good amount of power to the aggregate system.
Not only the GPU in mach2 is much more powerful than the one in mach1, but also it benefits from a double memory throughput.

In either case, \texttt{hgemms} fully adapts to the underlying hardware, properly exploiting its power.
Based on our results, we can confirm that \texttt{hgemms} is able to exploit ALP in CPU+GPU+XPU environments for the GEMM use case.
Therefore, this shows how the POAS framework can be applied to a specific domain and achieve potential performance enhancements.

\begin{table}[t]
\begin{tabular}{rllllll}
\toprule
      & \multicolumn{3}{c}{mach1}     & \multicolumn{3}{c}{mach2} \\ \cmidrule(l){2-4} \cmidrule(l){5-7}
Input & CPU    & GPU     & XPU     & CPU     & GPU     & XPU     \\ \midrule
   i1 & 0.32\% & 21.26\% & 78.42\% & 1.12\% & 27.03\% & 71.85\% \\
   i2 & 0.32\% & 22.29\% & 77.39\% & 1.22\% & 27.30\% & 71.48\% \\
   i3 & 0.33\% & 26.72\% & 72.95\% & 1.25\% & 30.93\% & 67.82\% \\
   i4 & 0.31\% & 25.27\% & 74.42\% & 0.95\% & 30.26\% & 68.79\% \\
   i5 & 0.31\% & 20.10\% & 79.59\% & 0.99\% & 25.53\% & 73.48\% \\
   i6 & 0.28\% & 22.77\% & 76.95\% & 1.07\% & 27.50\% & 71.43\% \\
\bottomrule
\end{tabular}
\caption{Percentage of work distribution among devices} \label{tab:perc}
\end{table}

\begin{figure}[t]
\includegraphics[width=0.94\linewidth]{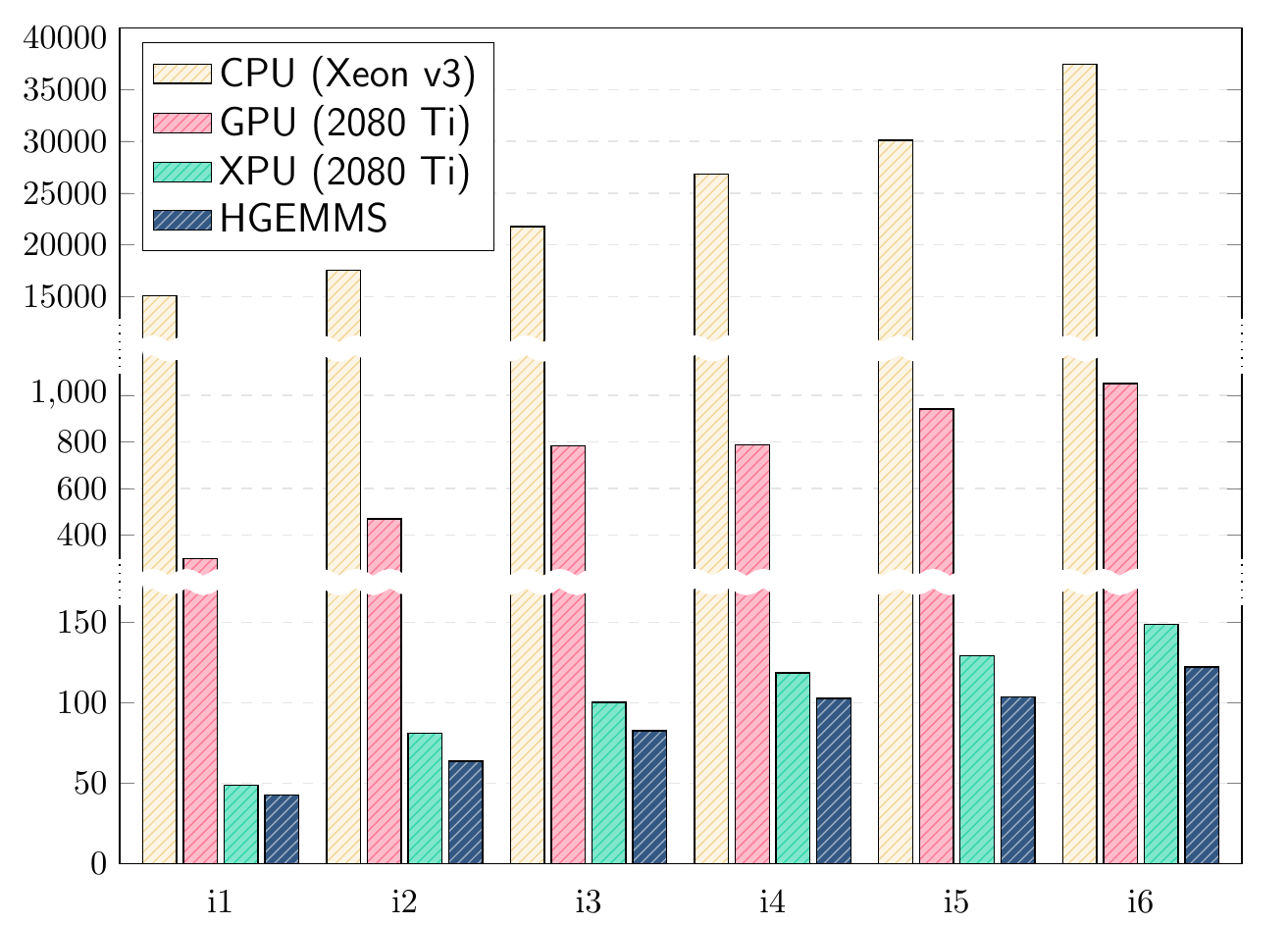}
\caption{Execution time of different input matrices in mach1} \label{fig:time}
\end{figure}
\begin{figure}[t]
\includegraphics[width=0.94\linewidth]{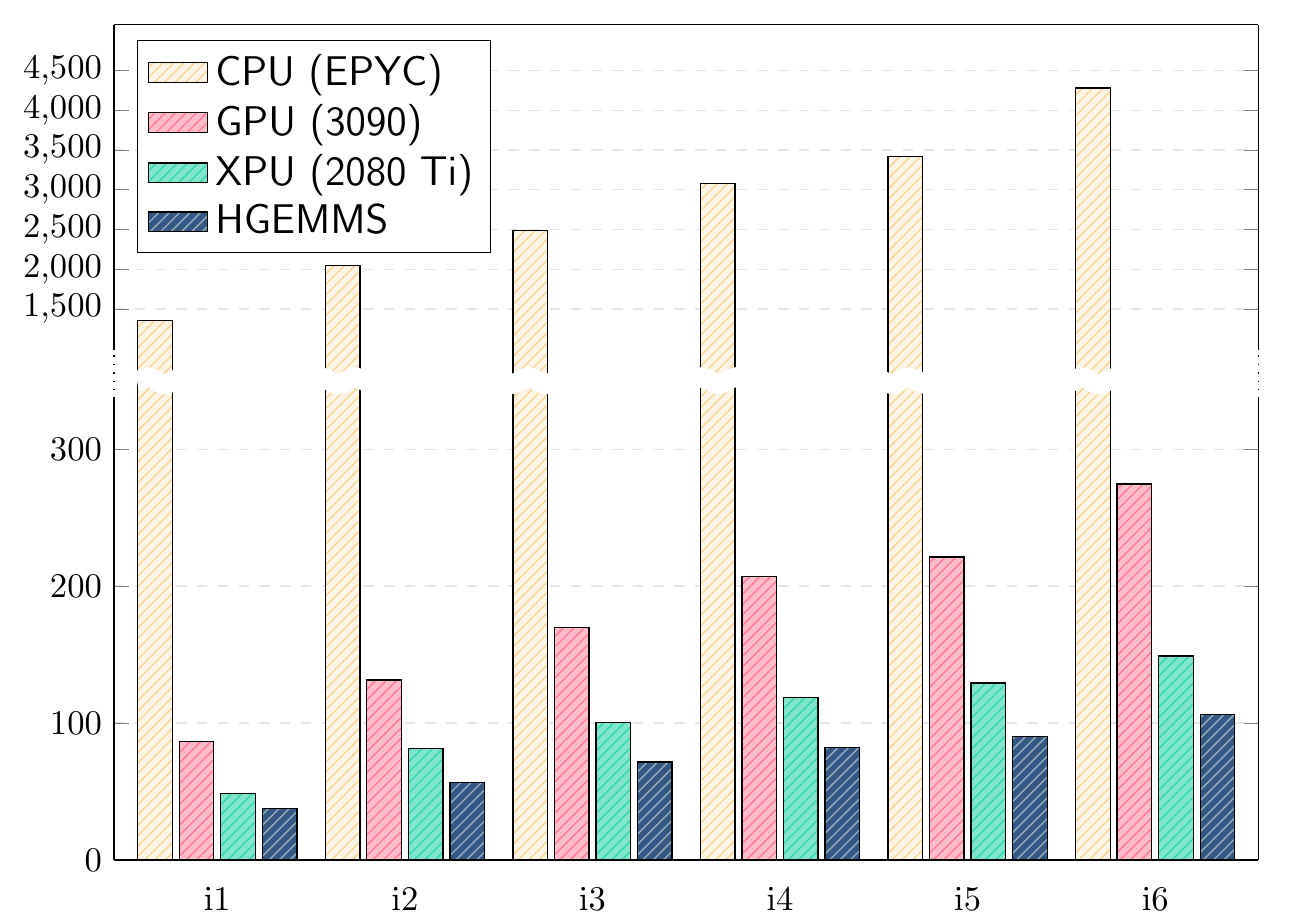}
\caption{Execution time of different input matrices in mach2} \label{fig:time2}
\end{figure}

\begin{table}[t]
\begin{tabular}{rllllll}
\toprule
   & \multicolumn{3}{c}{mach1} & \multicolumn{3}{c}{mach2} \\ \cmidrule(l){2-4} \cmidrule(l){5-7}
Input & CPU      & GPU   & XPU   & CPU     & GPU    & XPU    \\ \midrule
i1 & 353.02x & 7.04x & 1.14x & 35.93x & 2.30x & 1.29x \\
i2 & 275.89x & 7.39x & 1.28x & 36.21x & 2.32x & 1.43x \\
i3 & 263.72x & 9.49x & 1.22x & 34.71x & 2.37x & 1.40x \\
i4 & 261.09x & 7.66x & 1.15x & 37.51x & 2.52x & 1.45x \\
i5 & 290.23x & 9.07x & 1.25x & 37.86x & 2.45x & 1.43x \\
i6 & 306.72x & 8.60x & 1.22x & 40.20x & 2.58x & 1.40x \\
\bottomrule
\end{tabular}
\caption{Speedup of hgemms with respect to standalone execution} \label{tab:speedup}
\end{table}

\section{Conclusions and future work} \label{sec:conclusions}
Heterogeneity is becoming increasingly common in all markets.
Energy-constrained systems benefit from accelerators thanks to their lower consumption, while can also provide great performance in compute-intensive workloads.
One of the most promising views to exploit heterogeneity is Accelerator Level Parallelism (ALP).
However, to concurrently run any workload in multiple devices, the work has to be divided efficiently between the compute elements.
As the number of applications in which accelerators are being used grows rapidly, we need solutions that allow this process to be performed efficiently.

This work has presented POAS, a framework for defining scheduling models that efficiently divide and schedule any workload among the compute elements available within a node.
Our method adapts to the hardware being used, thanks to the performance prediction, thus maximizing resource usage.
We applied our framework to the field of matrix multiplication, showing that POAS adapts to the specific problem seamlessly.
Furthermore, our POAS application to GEMM showed excellent performance results in two HPC servers with a CPU+GPU+XPU configuration (for Intel and AMD CPUs).
Our implementation uses MKL and BLIS for CPU and cuBLAS for both GPU and XPU, achieving speedups of up to 45\% with respect to using only one accelerator.

For future work, we plan to further extend POAS with more sophisticated scheduling policies.
Although the proposed ones can provide high levels of hardware utilization in many scenarios, we believe that exploring new approaches can enhance the flexibility of the POAS framework to other domains.
Another research line that remains open is to provide POAS capabilities to determine when a given workload is appropriate for exploiting ALP or not.
Currently, the POAS framework can detect when running a certain workload is beneficial for co-execution or not depending on the amount of work to do.
However, it is only capable of doing so when the workload size is known (after the DS-POAS was designed), whereas applications that are unsuitable for co-execution could be detected prematurely.
Lastly, another front that remains open is how to efficiently schedule the communications between the CPU and accelerators.
This aspect can also have a notable impact on overall performance, and is something that must be carefully studied to further enhance the POAS framework.

\begin{acks}
Grant RTI2018-098156-B-C53 funded by MCIN/AEI/10.13039/501100011033 and by ``ERDF A way of making Europe''.
\end{acks}

\bibliographystyle{ACM-Reference-Format}
\bibliography{refs}

\end{document}